\documentclass[onecolumn,12pt]{revtex4}
\usepackage{amssymb}
\usepackage{latexsym}
\usepackage{epsfig}
\usepackage{amsmath}
\usepackage{graphicx}
\usepackage{epsfig}
\usepackage{bm}
\usepackage{cancel}
\usepackage{amssymb}
\usepackage{float}
\usepackage{subfigure}
\usepackage{dcolumn}
\usepackage[colorlinks]{hyperref}
\usepackage[usenames,dvipsnames]{color}

\hypersetup{
     breaklinks=true,
    pdfstartview={FitH},    
    colorlinks=true,       
    linkcolor=blue,          
    citecolor=red,        
    filecolor=magenta,      
    urlcolor=blue,           
    anchorcolor=green,      
    linktocpage=true
}
\begin{document}
\title{An extended analysis for a generalized Chaplygin gas model}
\author{Abdulla Al Mamon}
\email{abdulla.physics@gmail.com}
\affiliation{Department of Physics, Vivekananda Satavarshiki Mahavidyalaya (affiliated
to the Vidyasagar University), Manikpara-721513, West Bengal, India}
\author{Andronikos Paliathanasis}
\email{anpaliat@phys.uoa.gr}
\affiliation{Institute of Systems Science, Durban University of Technology, PO Box 1334,
Durban 4000, Republic of South Africa}
\affiliation{Instituto de Ciencias F\'{\i}sicas y Matem\'{a}ticas, Universidad Austral de
Chile, Valdivia 5090000, Chile}
\author{Subhajit Saha}
\email{subhajit1729@gmail.com}
\affiliation{Department of Mathematics, Panihati Mahavidyalaya, Kolkata 700110, West Bengal, India}
\begin{abstract}
In this work, we have extended the analysis on the generalized Chaplygin gas (GCG) model as the unification of dark energy and dark matter. Specifically,  we have shown that the model of our consideration known as the new generalized Chaplygin gas (NGCG) model, admits a scalar field description, which means that there exist a minimally coupled scalar field for a given scalar field potential where the equation of state is that of the NGCG. With the use of the later property we can construct the slow-roll parameters and derive the corresponding values for the spectral indices for the tensor to scalar perturbation and for the density perturbations. We have also studied the growth rate of matter perturbations in the NGCG scenario. Finally, we have studied the viability of the generalized second law of thermodynamics by assuming that the dynamical apparent horizon in a NGCG universe is endowed with Hawking temperature and Bekenstein entropy.
\end{abstract}
\maketitle
\section{Introduction}
 The currently observed accelerated expansion phase of the Universe is widely supported by different observational data \cite{acc1,acc2,acc3,acc4,acc5}. In the Einstein theory of gravity, dark energy (DE), a hypothetical exotic fluid, is presumed to be responsible for this cosmic acceleration and we refer the interested reader to Refs. \cite{de1,de2,de3} for a detailed description of DE. A host of observational data has been analyzed to deduce that only a small fraction, around 5$\%$ of the total energy density in the Universe, is in the form of baryonic matter, while dark matter (DM), which is required to explain structure formation, accounts for around 26$\%$ of the total energy density. The dominant component, DE, accounts for the remaining 69$\%$, which brings the total amount of energy density in the Universe close to the critical energy density \cite{p2015}. The simplest model for DE is the standard $\Lambda$CDM. The inclusion of cosmological constant $\Lambda$ with cold dark matter (CDM) brings the $\Lambda$CDM model into excellent agreement with the observed data. A cosmological constant is described by a single parameter, $\Lambda$, whose energy density remains constant with time and its equation of state (EoS) parameter is $\omega_{\Lambda}=-1$. Despite the great success of $\Lambda$CDM model, the cosmological constant suffers from the {\it fine-tuning problem} \cite{ftplcdm} and the {\it cosmic coincidence problem} \cite{ccplcdm}. For a recent discussion on the $\Lambda$CDM model, we refer the reader to Ref. \cite{leand}. In order to overcome these issues, various DE models has been explored in the literature, which include quintessence, K-essence, phantom, tachyon, and many more (for a review, one may see Ref. \cite{de1} and the references therein). However, we are yet to have a full-proof DE model.\\

\par As is well-known, the Chaplygin gas (CG) model as a unification of the unknown dark sectors (DM and DE) in the Universe is a good candidate amongst different DE models \cite{cgm,41}. In the CG model, the dark sectors can be unified by using an exotic equation of state (EoS) parameter. The striking property of this unified model is that the CG behaves as a dust-like matter (pressureless DM) at early times and behaves like a cosmological constant at late times. But, the CG models are under strong observational pressure from CMB anisotropies \cite{cgdb1,cgdb2}. This has led to the study of several interacting CG models in the literature \cite{Zhang1,Saha0}. It is worthwhile to mention here that Kamenshchik et al. \cite{cgm} also introduced a single-fluid model of a generalized chaplygin gas (GCG) with a modified equation of state $p=-\frac{A}{\rho^{\alpha}}$ where $A>0$ and $\alpha$ is a real number, and it reduces to CG when $\alpha=1$. This GCG model has been studied in some detail in Refs. \cite{gcg1,gcg2} and has been confronted with different phenomenological tests involving type Ia supernovae (SNIa), cosmic microwave background (CMB), and other observational datasets \cite{gcgobs1,gcgobs2,gcgobs3,gcgobs4,gcgobs5,gcgobs6}. Furthermore, it should be noted that the GCG model can be portrayed as an interacting $\Lambda$CDM model \cite{gcgobs6} in which a cosmological constant type DE interacts with cold DM. Later, Zhang et al. \cite{ngcg} proposed an extended version of the GCG which they termed it as the new generalized Chaplygin gas (NGCG) model, in order to describe the unification of DE and DM. They further demonstrated that the NGCG actually is a kind of interacting XCDM (X-matter with cold DM) model. Some recent studies on this new GCG and its applications in Cosmology can be found in \cite{ngcg,ngcg01,ngcg02,ngcg03,ngcg04,ngcg04a,mamonngcg}. \\

\par On the other hand, CG and GCG models play an important role for the description of inflation \cite{jd1,jd2,anin1,anin2}. During the inflationary period, the Universe is presumed to be dominated by a scalar field known as inflaton. In the slow-roll regime of the scalar field dynamics inflation occurs.  It has been found that various GCG models can be described by the scalar field dynamics with a slow-roll regime. Motivated by these facts, in the present work, we determine the scalar field equivalence for this model while we determine the slow-roll parameters and the spectral indices provided by the NGCG model. We also study the behavior of this model at the perturbation level as well. More specifically, we study the growth rate of matter perturbations in terms of evolution of the linear matter density contrast and explain how the growth rate data can be used to obtain predictions on the theoretical model under consideration. We also compare our results with the predictions of the standard $\Lambda$CDM and GCG models. Furthermore, we also study the viability of the generalized second law of thermodynamics in the NGCG model by assuming that the dynamical apparent horizon is endowed with Hawking temperature and Bekenstein entropy.\\

\par The rest of this paper is organized as follows. In the next section, the NGCG model as the unification of DE and DM is introduced briefly. In section \ref{sec-sfndinf}, we investigate the NGCG as a candidate for the description of inflation. We study the influence of this model on the structure formation in section \ref{sec-growth}. In section \ref{gsl}, we study the generalized second law of thermodynamics in the NGCG model at the dynamical apparent horizon with a view to garner support for the thermodynamic viability of the NGCG model. Finally, we present our conclusions in section \ref{conclusion}.\\
\par Throughout the paper, we use natural units such that $c=G=\hbar=1$.
\section{New generalized Chaplygin gas (NGCG) model}\label{sec2}
In what follows, we briefly introduce the NGCG model. We assume that the Universe is homogeneous and isotropic, and endowed with the spatially flat Friedmann-Lemaître-Robertson-Walker (FLRW) metric. The EoS of NGCG fluid is given by \cite{ngcg}
\begin{equation}\label{eq-rho-gen}
p_{NGCG}=-\frac{\tilde{A}(a)}{\rho^{\alpha}_{NGCG}},
\end{equation}
where $\alpha$ is the constant parameter, $\rho_{NGCG}$ is the energy density of the NGCG fluid, and the function $\tilde{A}(a)$ depends upon the scale factor, $a$, of the Universe. It might be expected that the NGCG fluid smoothly interpolates between a dust (pressureless matter) dominated phase ($\rho_{NGCG} \sim a^{-3}$ when $a$ is small) and a DE dominated phase ($\rho_{NGCG} \sim a^{-3(1+\omega_{de})}$ when $a$ is large), where $\omega_{de}$ is a constant EOS parameter. Also, the energy density of the NGCG fluid can be elegantly expressed as \cite{ngcg}
\begin{equation}\label{eqrhongcg1}
\rho_{NGCG}=[Aa^{-3(1+\omega_{de})(1+\alpha)} + Ba^{-3(1+\alpha)}]^{\frac{1}{1+\alpha}},
\end{equation}
where $A$ and $B$ are positive constants. It deserves to mention here that Eq. (\ref{eqrhongcg1}) can be deduced by substituting the EoS Eq. (\ref{eq-rho-gen}) into the energy conservation equation of the NGCG fluid. This requires the function $\tilde{A}(a)$ to be of the form
\begin{equation}
\tilde{A}(a)=-\omega_{de}Aa^{-3(1+\omega_{de})(1+\alpha)}.
\end{equation}
Finally, from Eq. (\ref{eqrhongcg1}) it follows that
\begin{equation}
\rho_{NGCG}=\rho_{NGCG0}a^{-3}[1-A_{s} + A_{s}a^{-3\omega_{de}(1+\alpha)}]^{\frac{1}{1+\alpha}},
\end{equation}
where the parameters $A$ and $B$ are redefined as $A_{s}=\frac{A}{A+B}$ and $\rho_{NGCG0}=(A+B)^{\frac{1}{1+\alpha}}$ denotes the present value of $\rho_{NGCG}$. For the present model, as a scenario of the unification of DE and DM, the energy densities of the DM and the DE components can be obtained, respectively, as \cite{ngcg}
\begin{equation}\label{eq-ngcgrdms1}
\rho_{dm}=\rho_{dm0} a^{-3} \times [1-A_{s} + A_{s}a^{-3\omega_{de}(1+\alpha)}]^{\frac{1}{1+\alpha} -1}
\end{equation}
and
\begin{equation}\label{eq-ngcgrdes1}
\rho_{de}=\rho_{de0} a^{-3[1+\omega_{de}(1+\alpha)]} \times [1-A_{s} + A_{s}a^{-3\omega_{de}(1+\alpha)}]^{\frac{1}{1+\alpha} -1},
\end{equation}
where $\rho_{dm0}$ and $\rho_{de0}$ are the current values of $\rho_{dm}$ and $\rho_{de}$, respectively. It can be readily observed that the NGCG model reduces to the standard $\Lambda$CDM model when we put $\alpha=0$ and $\omega_{de}=-1$. We also note that it becomes $\omega$CDM model when $\alpha=0$. It is also remarkable to observe that the GCG model is included as a sub-case, and can be obtained for $\omega_{de}=-1$. From Eqs. (\ref{eq-ngcgrdms1}) and (\ref{eq-ngcgrdes1}), one can now easily obtain the scaling behavior of the energy densities as
\begin{equation}
\frac{\rho_{dm}}{\rho_{de}}=\frac{\rho_{dm0}}{\rho_{de0}}a^{3\omega_{de}(1+\alpha)}.
\end{equation}
For $\alpha \neq 0$, there must exist an energy flow between the dark sectors (DE and DM) and thus, $\alpha$ describes the interaction between the dark sectors in this model \cite{ngcg}.\\

\par Here, we assume that the Universe is filled with NGCG fluid ($\rho_{NGCG}=\rho_{de}+\rho_{dm}$), baryonic matter component ($\rho_{b}$), and radiation component ($\rho_{r}$) so that the total energy density of the Universe can be obtained as $\rho_{total}=\rho_{NGCG}+\rho_{b}+\rho_{r}$. In the framework of a homogeneous, isotropic, and spatially flat FLRW cosmology, the Friedmann equation can be expressed as
\begin{equation}
3H^{2}=\rho_{total}.
\end{equation}
The dimensionless Friedmann equation $E(a)$ can be written as
\begin{eqnarray}\label{eqH}
E(a)= \frac{H(a)}{H_{0}}=[(1-\Omega_{b0}-\Omega_{r0})a^{-3} \times [1 - A_{s}  (1-a^{-3\omega_{de}\eta})]^{\frac{1}{\eta}} + \Omega_{b0}a^{-3} + \Omega_{r0}a^{-4} ]^{\frac{1}{2}},
\end{eqnarray}
where $\eta=(1+\alpha)$ and $H_{0}$ is the value of the Hubble parameter $H(a)$ at the present epoch. In the above equation, $\Omega_{b0}$ and $\Omega_{r0}$ denote the current values of dimensionless energy densities of baryonic matter and radiation, respectively.
\section{NGCG as inflationary model}\label{sec-sfndinf}
In this section, we investigate the NGCG (\ref{eq-rho-gen}) as candidate for
the description of the early acceleration phase of the Universe known as
inflation. Fluids with equation of state parameters which are
generalizations of the CG have been widely applied in the
literature as inflationary models see for instance \cite{jd1,jd2,anin1,anin2}
and references therein.\\

In order to perform such analysis we study if the NGCG model can be
described by a quintessence scalar field such that the slow-roll description
for the inflation to exist. Then, we define the slow-roll parameters and we
calculate the values for the spectral indices and we compare with the
observations as they are provided by Planck collaboration.

\subsection{Scalar field description}

Assume now a spatially flat FLRW spacetime where the matter source is
described by a homogeneous scalar field minimally coupled to gravity \cite{qq}, the inflaton field, such that the Einstein's field equations to be
derived%
\begin{equation}
3H^{2}=\frac{1}{2}\dot{\phi}^{2}+V(\phi ),  \label{in.03}
\end{equation}%
\begin{equation}
2\dot{H}+3H^{2}=-\frac{1}{2}\dot{\phi}^{2}+V(\phi ),  \label{in.04}
\end{equation}%
where $\rho _{\phi }=\frac{1}{2}\dot{\phi}^{2}+V(\phi )$, $p_{\phi }=\frac{1%
}{2}\dot{\phi}^{2}-V(\phi )$ and the continuous equation $\dot{\rho}_{\phi
}+3H\left( \rho _{\phi }+p_{\phi }\right) =0$ reads
\begin{equation}
\ddot{\phi}+3H\dot{\phi}+V_{,\phi }=0.  \label{in.05}
\end{equation}

Function $V\left( \phi \right) $ is the scalar field potential and defines
the evolution of the scalar field and the description of the inflationary
period, for some examples we refer the reader in \cite{inf1,inf2,inf3}.\\

We follow the analysis presented in \cite{anin1} where we are able to write
the algebraic solution for the field equations (\ref{in.03})-(\ref{in.05}).
Indeed, in the new change of variables $dt=\exp \left( \frac{F\left( \chi
\right) }{2}\right) d\chi $ with $\chi =6\ln a,$ the background space reads%
\begin{equation}
ds^{2}=-e^{F\left( \chi \right) }d\chi ^{2}+e^{\chi
/3}(dx^{2}+dy^{2}+dz^{2}).  \label{in.14}
\end{equation}%
where the scalar field $\phi \left( \chi \right) $ and the scalar field
potential $V\left( \phi \left( \chi \right) \right) $ are given by the
formulas%
\begin{equation}
\phi (\chi )=\pm \frac{\sqrt{6}}{6}\int \!\!\sqrt{F^{\prime }(\chi )}d\chi ~,
\label{in.12}
\end{equation}%
\begin{equation}
V(\chi )=\frac{1}{12}e^{-F(\chi )}\left( 1-F^{\prime }(\chi )\right) ,
\label{in.13}
\end{equation}%
in which a prime \textquotedblleft $^{\prime }$\textquotedblright\ denotes
total derivative with respect to $\chi $, i.e. $F^{\prime }\left( \chi
\right) =\frac{d}{d\chi }F\left( \chi \right) $. The corresponding Hubble
function is defined as $H\left( \chi \right) =\frac{1}{6}\sqrt{e^{-F\left(
x\right) }}$.\\

In the new variables the energy density and pressure component are expressed
$\rho _{\phi }(\chi )=\frac{1}{12}e^{-F(\chi )}~$,$~p_{\phi }(\chi )=\frac{1%
}{12}e^{-F(\chi )}\left( 2F^{\prime }(\chi )-1\right) $. Furthermore, the
EoS parameter is defined $w_{eff}\left( \chi \right) =\left(
2F^{\prime }(\chi )-1\right) .$

Hence, from the EoS parameter (\ref{eq-rho-gen}) we define the
ordinary differential equation%
\begin{equation}
\frac{1}{12}e^{-F(\chi )}\left( 2F^{\prime }(\chi )-1\right) =\omega
_{de}Ae^{\chi -\frac{2}{2}(1+\omega _{de})(1+\alpha )}\left( \frac{1}{12}%
e^{-F(\chi )}\right) ^{\alpha }
\end{equation}%
with analytic solution%
\begin{equation}
F\left( \chi \right) =\frac{1+\omega _{de}}{2}\chi -\frac{1}{1+\alpha }\ln
\left( \bar{A}+\exp \left( \frac{\left( 1+\alpha \right) \omega _{de}}{2}%
\chi \right) \right) ,~\alpha \neq -1.  \label{ss1}
\end{equation}%
where $\bar{A}=12^{-\left( 1+\alpha \right) }A$. \ Consequently, for the
scalar field we calculate%
\begin{equation}
\phi \left( \chi \right) =\frac{1}{2\sqrt{3}\left( 1+\alpha \right) \omega
_{de}}\left( \left( 1+\alpha \right) \omega _{de}\sqrt{1+\omega _{de}}\chi -2%
\sqrt{1+\omega }\ln \left( 2\bar{A}\left( 1+\omega _{de}\right) +\phi
_{1}\left( \chi \right) \right) +2\ln \left( \phi _{2}\left( \chi \right)
\right) \right) ~,
\end{equation}%
\begin{equation}
V\left( \chi \right) =\frac{1}{24}e^{-\frac{1+\omega _{de}}{2}x}\left( \bar{A%
}+e^{\frac{1+\omega _{de}}{2}x}\right) ^{-1+\frac{1}{1+\alpha }}\left( \bar{A%
}\left( 1-\omega _{de}\right) +e^{\frac{1+\omega _{de}}{2}x}\right) ~,
\end{equation}%
with%
\begin{equation}
\phi _{1}\left( \chi \right) =\left( 2+\omega _{de}\right) e^{\frac{1+\omega
_{de}}{2}x}+2\sqrt{\left( 1+\omega _{de}\right) \left( \bar{A}+e^{\frac{%
1+\omega _{de}}{2}x}\right) \left( \bar{A}\left( 1+\omega _{de}\right) +e^{%
\frac{1+\omega _{de}}{2}x}\right) }~,
\end{equation}%
\begin{equation}
\phi _{2}\left( \chi \right) =\left( 2+\omega _{de}\right) \bar{A}+2\left(
e^{\frac{1+\omega _{de}}{2}x}+\sqrt{\left( \bar{A}+e^{\frac{1+\omega _{de}}{2%
}x}\right) \left( \bar{A}\left( 1+\omega _{de}\right) +e^{\frac{1+\omega
_{de}}{2}x}\right) }\right) .
\end{equation}

\begin{figure}[tbp]
\centering\includegraphics[width=1\textwidth]{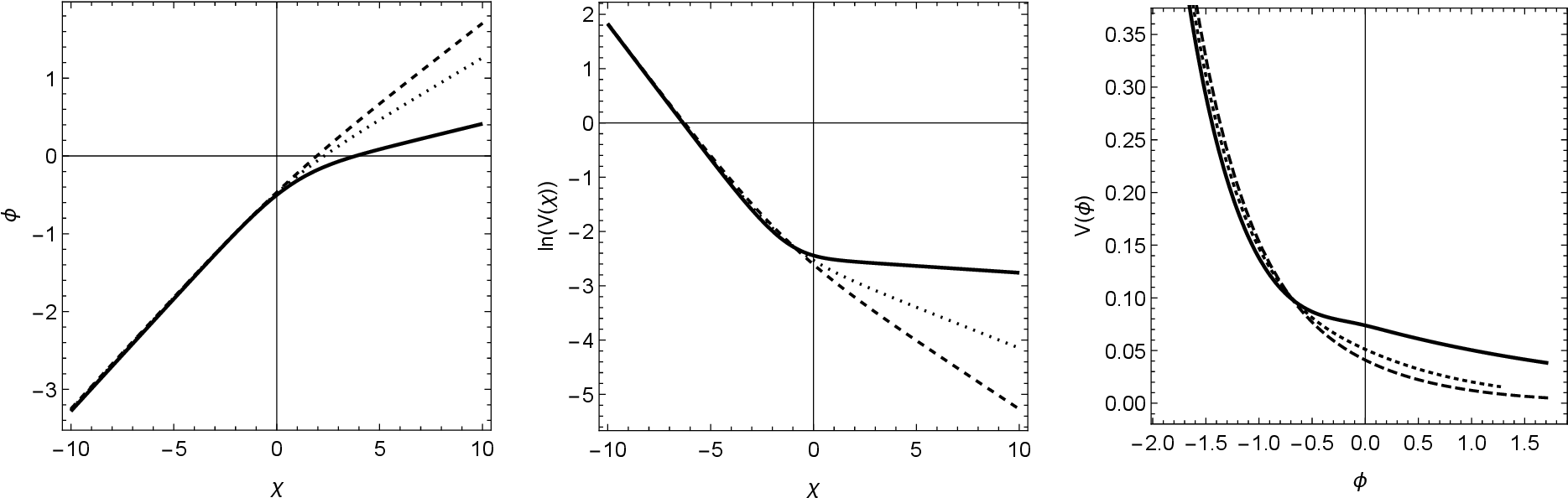}
\caption{Qualitative evolution for the scalar field $\protect\phi \left(
\protect\chi \right) $ (left Fig.) and the scalar field potential $V\left(
\protect\chi \right) $ (middle Fig.). Plots are for $\bar{A}=1$ and $\protect%
\alpha =1$. Solid lines are for $\protect\omega _{de}=-0.95,$ dotted lines
are for $\protect\omega _{de}=-0.7$ and dashed lines are for $\protect\omega %
_{de}=-0.5$. Right figure is the parametric plot $\protect\phi \left(
\protect\chi \right) -V\left( \protect\phi \left( \protect\chi \right)
\right) $.}
\label{fig1}
\end{figure}

\begin{figure}[tbp]
\centering\includegraphics[width=0.32\textwidth]{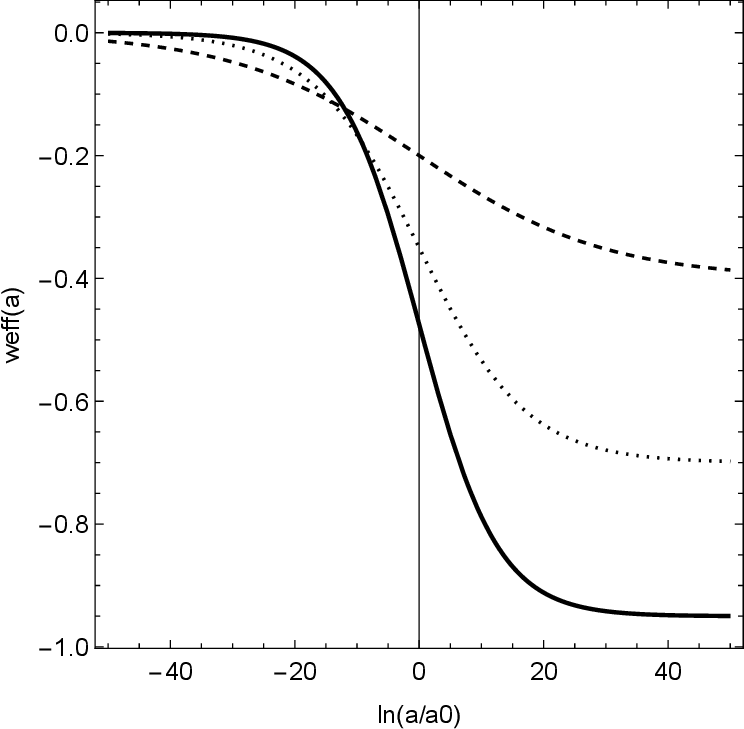}
\caption{Qualitative evolution for the equation of state evolution $%
w_{eff}\left( a\right) $ for the NGCG of our consideration. Plot is for $%
\bar{A}=1$ and $\protect\alpha =1$. Solid line is for $\protect\omega %
_{de}=-0.95,$ dotted line is for $\protect\omega _{de}=-0.7$ and dashed line
is for $\protect\omega _{de}=-0.5$.}
\label{fig2}
\end{figure}

In Fig. \ref{fig1} we present the qualitative evolution for the scalar field
which is equivalent for the NGCG of our consideration. We remark that there
exists a real scalar field and a positive valued scalar field potential
which provides the same evolution for the background space as the NGCG
model. Furthermore, in Fig. \ref{fig2} \ we present the qualitative
evolution for the equation of state parameter $w_{eff}\left( a\right) $.
\qquad \qquad

\subsection{Slow-roll parameters and spectral indices}

In the inflationary era, the scalar field potential dominates such that $%
3H^{2}\simeq V\left( \phi \right) $, while $\dot{\phi}\simeq -\frac{V_{,\phi
}}{3H}$. The a specific scalar field potential, the following parameters are
defined \cite{l01}
\begin{equation}
\varepsilon _{V}=\left( \frac{V_{,\phi }}{2V}\right) ^{2}~\,,~\eta _{V}=%
\frac{V_{,\phi \phi }}{2V},  \label{in.08}
\end{equation}%
which are known as potential slow-rolls parameters. The condition for
inflation is $\varepsilon _{V}<<1$, while the additional condition is
introduced $\eta _{V}<<1$, such that the  inflationary phase to last long
enough.\\

In a similar way, for a specific Hubble function the Hubble slow-roll
parameters are defined as follows \cite{l01} $\varepsilon _{H}=-\frac{d\ln H%
}{d\ln a}$ and $\eta _{H}=-\frac{d\ln H_{,\phi }}{d\ln a}\,$. The two
different set of parameters are related as follows $\varepsilon _{V}\simeq
\varepsilon _{H}~$and~$\eta _{V}\simeq \varepsilon _{H}+\eta _{H}.$%
Consequently, as in the case of the potential slow-roll parameters, the
inflationary era is recovered when $\varepsilon _{H}<<1$, while $\eta _{H}<<1
$ is the second condition. Because, we do not have a closed-form expression
for the scalar field potential we select to work with the Hubble slow-roll
parameters.\\

The slow-roll parameters are related with the spectral indices for the
density perturbations $n_{s}$, and the tensor to scalar ration $r$ as follow%
\begin{equation}
r=10\varepsilon _{H}~,
\end{equation}%
\begin{equation}
n_{s}=1-4\varepsilon _{H}+2\eta _{H}.
\end{equation}

From the analysis of the cosmological observations of the Planck 2018
collaboration \cite{pin2018}, the spectral index for the density
perturbations is constraint as $n_{s}=0.9649\pm 0.0042$, while the tensor to
scalar ratio, $r$ is constraint as $r<0.10$.\\

Moreover, we can define the number of e-folds $N_{e}\left( \chi \right) $,
by the expression $N_{e}=\int_{t_{i}}^{t_{f}}H\left( t\right) dt=\ln \frac{%
a_{f}}{a_{i}}=\frac{1}{6}\left( \chi _{f}-\chi _{i}\right) ,$where $\chi _{f}
$ is the moment where the inflation ends, that is, $\varepsilon _{H}\left(
\chi _{f}\right) =1$.\\

Thus, the slow-roll parameters $r$ and $n_{s}$ in the first-order
approximation are expressed in term of the number of e-folds as follow%
\begin{equation}
r\left( N_{e}\right) =15\left( 1-\frac{\omega _{de}}{e^{3\left( 1+\alpha
\right) N_{e}\omega _{de}}-\left( 1+\omega _{de}\right) }\right)
\end{equation}%
\begin{equation}
n_{s}\left( N_{e}\right) =-2+\frac{3\left( 2+\alpha \right) \omega _{de}}{%
e^{3\left( 1+\alpha \right) N_{e}\omega _{de}}-1}-\frac{3\left( 1+\alpha
\right) \omega _{de}\left( 1+\omega _{de}\right) }{\left( 1+3\omega
_{de}\right) e^{3\left( 1+\alpha \right) N_{e}\omega _{de}}-\left( 1+\omega
_{de}\right) }\text{.}
\end{equation}

We observe that $r\left( N_{e}\right) $ and $n_{s}\left( N_{e}\right) $ are
sigmoid functions, where $\left( 1+\alpha \right) \omega _{de}<0$ and for
large values of, such as $N_{e}\simeq 55$ \cite{pin2018}, it follows $%
r\simeq 15\left( 1+\omega _{de}\right) $ and  $n_{s}\simeq -2-3\omega _{de}$
\ Hence, $\omega _{de}<-0.9933$, where then $n_{s}\simeq 1$. Hence, the NGCG
fits the observations when $\omega _{de}$ is near to the cosmological
constant term. It is important to mention that we have assumed that $\alpha $
is not close to $-1$. Because of the nature of the sigmoid function the
behaviour is different when$~1+\alpha \simeq 0$. However, such case is
excluded from definition of the model.\\

Thus, for large values of $N_{e}$, it follows that $r\left( n_{s}\right)
=5\left( 1-n_{s}\right) $, where for $n_{s}\rightarrow 1$, we find $%
r\rightarrow 0$, which is in agreement with the observations.
\section{Growth of matter perturbations in the NGCG model}\label{sec-growth}
In this section, we study the linear growth of DM fluctuations for the NGCG model. Assuming that the Universe is homogeneous and isotropic and that it can be described by either General Relativity or some modified gravity theory, we can study its consequences at the perturbations level by employing the effective Newtonian constant \cite{meq1,meq2,meq3,meq4,meq5,meq6,meq7,meq8,meq9,meq10,giapr1,giap1,giap2}. Using the subhorizon approximation ($k>>aH$), it can be shown that the linear matter perturbations grow according to the following second-order differential equation
\cite{meq1,meq2,meq3,meq4,meq5,meq6,meq7,meq8,meq9,meq10}
\begin{equation}\label{eq-deltam}
\delta^{\prime\prime}(a)+\left(\frac{3}{a}+\frac{H^{\prime}(a)}{H(a)}\right) - \frac{3}{2}\frac{\Omega_{m}(a) G_{eff}(a,k)/G_{N}}{a^2}\delta (a) =0,
\end{equation}
where primes denote differentiation with respect to the scale factor and $G_{eff}$ is the effective Newton's constant which is equal to $G_{N}$ in General Relativity, while, in general modified gravity models, $G_{eff}$ can be dependent on the scale factor, $a$, and the wave number, $k$, of the modes of the perturbations in Fourier space. It is important to note that on subhorizon scales ($k>>aH$), we may ignore the dependence on the wave number $k$ for both $G_{eff}$ and $\delta$ \cite{kahsub}. In Eq. (\ref{eq-deltam}), $\delta=\frac{\delta \rho_{m}}{\rho_{m}}$ represents the growth function of the linear matter density contrast in which $\rho_{m}$ denotes the background matter density and $\delta \rho_{m}$ denotes its first-order perturbation. It is worthwhile to mention here that, for the present case,  $G_{eff}=G_{N}$ and $\Omega_{m}(a)=\Omega_{b}(a)+\Omega_{dm}(a)$. The corresponding expressions for $\Omega_{b}(a)$ and $\Omega_{dm}(a)$ are given by
\begin{equation}\label{eq-ombb}
\Omega_{b}(a)=E^{-2}(a) \times \Omega_{b0}a^{-3}
\end{equation}
and
\begin{equation}\label{eq-omdmm}
  \Omega_{dm}(a)=E^{-2}(a) \times \Omega_{dm0}a^{-3} [1-A_{s} + A_{s}a^{-3\omega_{de}\eta}]^{\frac{1}{\eta} -1}
\end{equation}
respectively. Since we will discuss on low redshifts, we ignore the radiation component, i.e. $\Omega_{r0}=0$. In that case, Eq. (\ref{eqH}) becomes
\begin{eqnarray}\label{eqH2.0}
E(a)= \frac{H(a)}{H_{0}}=[(1-\Omega_{b0})a^{-3} \times [1 - A_{s}  (1-a^{-3\omega_{de}\eta})]^{\frac{1}{\eta}} + \Omega_{b0}a^{-3}]^{\frac{1}{2}}.
\end{eqnarray}
\par
\begin{figure}[tbp]
\centering
\includegraphics[width=0.45\textwidth]{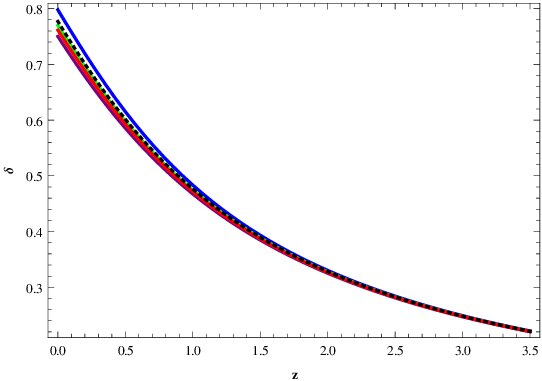}\hspace{7mm} \includegraphics[width=0.45\textwidth]{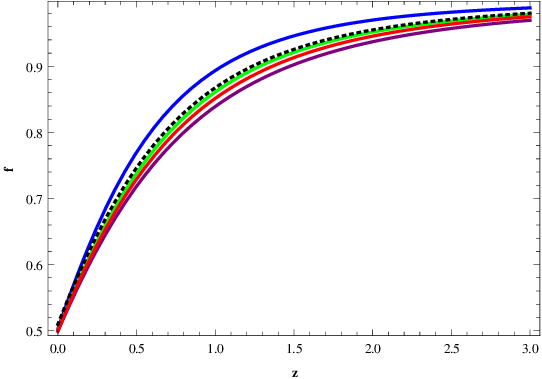}
\caption{The evolutionary trajectories of $\delta $ (left panel) and $f$ (right panel) are shown as a function of redshift $z (\equiv \frac{1}{a}-1)$ for the NGCG model by considering the values of $\Omega_{b0}=0.046, \Omega_{dm0}=0.2508, A_{s}=0.7373$, and $\eta=0.9443$. The blue, green, and purple curves are for $\omega_{de}=-1.2$, $\omega_{de}=-1.041$, and $\omega_{de}=-0.95$, respectively. In each plot, the black and red curves represent the corresponding evolutions of $\delta (z)$ and $f(z)$ in $\Lambda$CDM ($\omega_{de}=-1, \eta=1$) and GCG ($\omega_{de}=-1$) models, respectively.}
\label{fig-da}
\end{figure}

\begin{figure}[tbp]
\centering\includegraphics[width=0.4\textwidth]{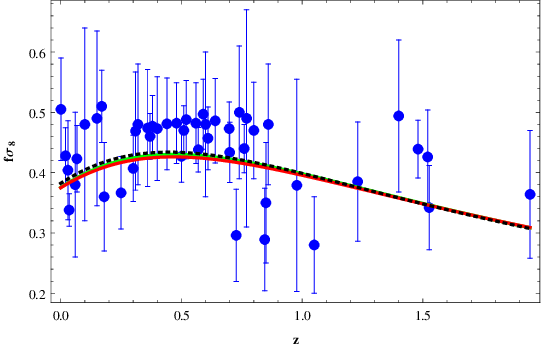}
\includegraphics[width=0.4\textwidth]{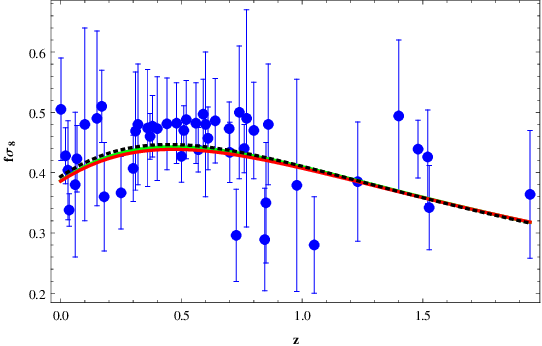}
\includegraphics[width=0.4\textwidth]{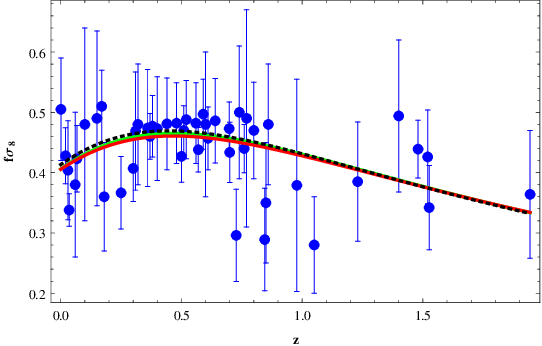}
\caption{The evolutionary trajectory of $f \sigma_{8}$, as a function of redshift $z$, is shown for the present model (green curve) by considering the values of $\Omega_{b0}=0.046, \Omega_{dm0}=0.2508, \omega_{de}=-1.041, A_{s}=0.7373$, and $\eta=0.9443$. The plots are for $\sigma_{8,0}=0.75$ \cite{s875} (left panel), $\sigma_{8,0}=0.772$ \cite{s8772} (right panel), and $\sigma_{8,0}=0.811$ \cite{s8811} (lower panel), respectively. In each plot, the black and red curves represent the corresponding evolutions of $f \sigma_{8}$ in a $\Lambda$CDM and GCG models, respectively. Also, the blue dots (with the error bars) correspond to the 47 Growth of matter data points (redshift space distortions measurements) in the redshift range $0.001 \le z \le 1.944$, obtained from different surveys and the corresponding $f \sigma_{8}$ values are given in \cite{refwx} and the relevant references therein.}
\label{fig-fs47}
\end{figure}

 Next, in order to study the influence of the NGCG model on the structure formation, we now discuss the growth rate $f$ and the root mean square (RMS) normalization of the matter power spectrum $\sigma_{8}$ of the NGCG model. The growth rate can be measured by the peculiar velocities of galaxies falling towards overdense regions and is defined as \cite{dgrf1,dgrf2,dgrf3}
\begin{equation}
f(a)={\rm \frac{d log ~\delta}{d log ~a}}.
\end{equation}
Also, the RMS fluctuations of the linear density field within a sphere of radius $R = 8h^{-1}~Mpc$ is defined as
\begin{equation}
\sigma_{8}(a)= \sigma_{8,0} \frac{\delta(a)}{\delta(1)},
\end{equation}
where $\sigma_{8,0}$ is the present value of $\sigma_{8}(a)$. However, a more robust and bias-free quantity that is measured by redshift surveys is the combination of $f (a)$ and $\sigma_{8} (a)$. Thus, we have
\begin{equation}\label{eq-fs-th}
f\sigma_{8}(a)=a \frac{\delta^{\prime} (a)}{\delta(1)} \sigma_{8,0}.
\end{equation}
This observable combination is measured at different redshifts by various cosmological surveys as a probe of the growth of matter density perturbations. Also, the theoretically predicted value of this combination can be evaluated from the solution, $\delta (a)$, of the Eq. (\ref{eq-deltam}) for a given set of cosmological parameters. Furthermore, one can now compare this theoretical prediction with the observed $f\sigma_{8}$ dataset. It is evident from Eqs. (\ref{eq-deltam}), (\ref{eq-ombb}) and (\ref{eq-omdmm})  that $\delta (a)$ depends on the functional form of $H(a)$ and will yield a different result when a different $H(a)$ is chosen. Hence, one can easily solve Eq. (\ref{eq-deltam}) to determine the evolution of $\delta(a)$ for a given parametrization of $H(a)$ with certain initial conditions. For the growing mode, we numerically solve the Eq. (\ref{eq-deltam}) so that $\delta(a_{i})=a_{i}$ and $\delta^{\prime}(a_{i})=1$, where $a_{i}$ is the initial scale factor which we have chosen to be $10^{-3}$. Moreover, we also use in our analysis the best fit values of the free parameters ($\Omega_{b0}, \Omega_{dm0}, \omega_{de}, A_{s}, \eta$) = ($0.046, 0.2508, -1.041, 0.7371, 0.9443$) obtained by Salahedin et al. \cite{ngcg01} using  $H(z)$ + BAO + CMB + BBN + SNIa (Pantheon) data. In this work, we have chosen three values of $\sigma_{8,0}$ given by $\sigma_{8,0}=0.75$ \cite{s875}, $\sigma_{8,0}=0.772$ \cite{s8772}, and $\sigma_{8,0}=0.811$ \cite{s8811}. Finally, in Fig. \ref{fig-da}, we have shown the evolutions of $\delta(z)$ and $f(z)$, as a function of the redshift $z$ $(\equiv \frac{1}{a}-1)$, for the NGCG model. The plot of $\delta$ as a function of the redshift parameter $z$ is shown in the left panel of Fig. \ref{fig-da}, while the corresponding plot of $f(z)$ is shown in the right panel of Fig. \ref{fig-da}. For comparison, the evolutions of $\delta(z)$ and $f(z)$ for a flat $\Lambda$CDM ($\omega_{de}=-1, \eta=1$) and GCG ($\omega_{de}=-1$) models are also shown. By comparing the evolutionary trajectories in Fig. \ref{fig-da}, it has been found that at early times, the $\delta$ and $f$ of these three models are the same, but the deviation among these models grows at late times. Furthermore, in Fig. \ref{fig-fs47}, we have compared the observed $f(z) \sigma_{8}(z)$ with the theoretical value of the growth rate function for this model. It is observed from Fig. \ref{fig-fs47} that the present model reproduces the observed values of $f(z) \sigma_{8}(z)$ quite effectively.
\section{Generalized Second Law of Thermodynamics in the NGCG Model} \label{gsl}
In this section, we wish to study the generalized second law in the NGCG model by considering the dynamical apparent horizon as the thermodynamic boundary. It is worthwhile to mention here that the generalized second law was adapted in the context of Cosmology by Brustein \cite{gsl-a}. This second law is based on the conjecture that causal boundaries and not only black hole event horizons have geometric entropies proportional to their area. The dynamical apparent horizon always exists irrespective of the cosmological model we choose. This property allows it to be a natural choice for the study of gravitational thermodynamics. The results obtained within this framework can then be extended to the whole Universe, thanks to the cosmological principle.\\

It must be noted that the dynamical apparent horizon is a thermodynamic entity which must be endowed with an entropy and a temperature. In analogy with a black hole event horizon in the study of black hole thermodynamics \cite{gsl-b,gsl-c}, the temperature associated with the dynamical apparent horizon is identified as the Hawking temperature whose expression is given by
\begin{equation} \label{haw-temp}
T_{A_{X}}=\frac{1}{2\pi R_{A_{X}}}\left(1-\frac{\dot{R}_{A_{X}}}{2}\right),
\end{equation}
where $R_{A_{X}}$ is the proper radius of the apparent horizon in the NGCG model. In the rest of this section, the suffix `$X$' will denote corresponding parameters in a NGCG-dominated universe. One should note that Eq. (\ref{haw-temp}) represents the non-truncated version of the Hawking temperature. It has been customary to use the truncated expression \cite{gsl01}
\begin{equation} \label{haw-temp-tr}
T_{A_{X}}^{(tr)}=\frac{1}{2\pi R_{A_{X}}}
\end{equation}
in the literature for the study of gravitational thermodynamics, however, Bin{\'e}truy and Helou \cite{gsl02,gsl03} have put forward several strong arguments against the use of the truncated Hawking temperature. Furthermore, the consideration of the non-truncated form of the Hawking temperature has presented us with some promising results in a recent paper \cite{gsl04} by one of the authors.\\

The entropy on the horizon is given by the Bekenstein entropy which has the expression \cite{gsl05}
\begin{eqnarray} \label{bek-ent}
S_{A_{X}} &=& \frac{A_{A_{X}}}{4} \nonumber \\
&=& \pi R_{A_{X}}^{2},
\end{eqnarray}
where $A_{A_{X}}=4\pi R_{A_{X}}^{2}$ is the proper area bounded by the dynamical apparent horizon.\\

Next, we require the time-derivative of the entropy of the cosmic fluid inside the apparent horizon, which is obtained from the Clausius relation
\begin{equation} \label{claus-rel}
T_{fA_{X}}dS_{fA_{X}}=dU+p_{NGCG}dV_{A_{X}},
\end{equation}
where, $S_{fA_{X}}$ and $T_{fA_{X}}$ are the entropy and the temperature of the cosmic fluid within the horizon, respectively, while $U=\frac{4}{3}\pi R_{A_{X}}^{3}\rho _{NGCG}$ is the internal energy of the fluid evaluated at the apparent horizon. We shall assume that the temperature of the cosmic fluid, $T_{fA_{X}}$, is equal to $T_{A_{X}}$ in Eq. (\ref{haw-temp}). This assumption is supported by the results obtained in the pioneering work by Mimoso and Pav{\'o}n \cite{gsl06}.\\

Now, using Eqs. (\ref{haw-temp}), (\ref{bek-ent}), and (\ref{claus-rel}), we obtain the time-derivative of the total entropy
\begin{equation}
\dot{S}_{A_{X}}=18 \pi R_{A_{X}}\frac{(1+w_{NGCG})^2}{(1-3w_{NGCG})},
\end{equation}
which directly follows from Eq. (19) of Ref. \cite{gsl04}. The generalized second law is nothing but the requirement that $\dot{S}_{A_{X}}$ be non-negative. In our model, this requirement is satisfied if $w_{NGCG} \leq \frac{1}{3}$. It is quite easy to see that the effective EoS of the NGCG model is given by
\begin{eqnarray}
w_{NGCG} &=& \frac{p_{NGCG}}{\rho_{NGCG}} \nonumber \\
&=& \frac{w_{de}}{1+\left(\frac{B}{A}\right)a^{3w_{de}(1+\alpha)}}.
\end{eqnarray}
Since the constants $A$ and $B$ are both positive, and $w_{de}$ is always negative, it follows that $w_{NGCG} \leq \frac{1}{3}$ is always true. Thus, the generalized second law is unconditionally valid in the NGCG model and consequently, this model is consistent with thermodynamics.

\section{Conclusions}\label{conclusion}
Cosmological fluids with equation of state similar to that of the Chaplygin Gas have been widely used for the study of the inflation. Moreover, they have been applied for the description of the inflaton field. In this study, we show that the model of our consideration, the NGCG model, admits a scalar field description, which means that there exist a minimally coupled scalar field for a given scalar field potential where the equation of state parameter is that of the NGCG. Hence, we were able to determine the slow-roll parameters for the NGCG and determine the spectral indices and compare their values with the observational ones. In the first-order approximation, we derived the closed-form expressions for the scalar to tensor ratio and for the spectral index of the density perturbations. These two indices were found to be functions of the number of e-folds, $N_{e}$, and of the two parameters $\alpha$ and $\omega_{de}$ of the NGCG model. In addition, the functional form is that of a sigmoid where we found that for $(1+\alpha) \omega_{de}<0$ and for large values of the number of e-folds, the spectral indices are related by the linear expression $r(n_s)=5(1-n_s)$. The later expression provides values for the spectral indices in agreement with the observations. \\

\par Furthermore, we have studied the growth of matter perturbations in the NGCG scenario. In particular, we have shown the evolutions of the matter density contrast $\delta(z)$ and the growth rate $f(z)$ for this model and compared it with that of the standard $\Lambda$CDM and the GCG models. By comparing these models (NGCG, $\Lambda$CDM, and GCG), we have shown that all of these three models are the same at early times, but the evolutionary trajectory of the NGCG model deviates from the other two at late times. We have also computed the combination parameter $f\sigma_{8}$ (\ref{eq-fs-th}) as a function of the redshift for this model. The numerical results are summarized in Fig. \ref{fig-fs47}, where the comparison with the redshift space distortions measurements is shown as well.\\

\par Finally, we have shown that the NGCG model is consistent with gravitational thermodynamics, a result which puts the NGCG model on a firm theoretical ground.
\section{Acknowledgements}
The authors are thankful to the anonymous referee whose useful suggestions have
improved the quality of the paper.


\begin{thebibliography}{99}
\bibitem{acc1}A. G. Riess et al., ``Observational evidence from supernovae for an accelerating universe and a cosmological constant,'' Astron. J. {\bf 116}, 1009 (1998).

\bibitem{acc2}S. Perlmutter et al., ``Measurements of $\Omega$ and $\Lambda$ from 42 high redshift supernovae,'' Astrophys. J. {\bf 517}, 565 (1999).

\bibitem{acc3}P. de Bernardis et al., ``A Flat universe from high resolution
maps of the cosmic microwave background radiation,'' Nature {\bf 404}, 955 (2000).

\bibitem{acc4}W. J. Percival et al., ``The 2dF Galaxy Redshift Survey: The Power
spectrum and the matter content of the Universe,'' Mon. Not. Roy. Astron. Soc. {\bf 327},
1297 (2001).

\bibitem{acc5}D. N. Spergel et al., ``First year Wilkinson Microwave Anisotropy
Probe (WMAP) observations: Determination of cosmological parameters,'' Astrophys.
J. Suppl. {\bf 148},  175 (2003).

\bibitem{de1} L. Amendola and S. Tsujikawa, {\it Dark Energy: Theory and Observations}, Cambridge
University Press, Cambridge, UK, (2010).

\bibitem{de2}K. Bamba, S. Capozziello, S. Nojiri, S. D. Odintsov, ``Dark energy cosmology: The equivalent description via different theoretical models and cosmography tests,'' Astrophys. Space Sci. {\bf 342}, 155 (2012).

\bibitem{de3}E. J. Copeland, M. Sami, S. Tsujikawa, ``Dynamics of dark energy,'' Int. J. Mod. Phys. D. {\bf 15}, 1753 (2006).

\bibitem{p2015}P. A. R. Ade et al. [Planck Collaboration], `` Planck 2015 results XIV. Dark energy and modified gravity,'' A$\&$A, {\bf 594}, A14 (2016).

\bibitem{ftplcdm}S. Weinberg, ``The cosmological constant problem,'' Rev. Mod. Phys. {\bf 61}, 1 (1989).

\bibitem{ccplcdm}P. J. Steinhardt, L. Wang, I. Zlatev, ``Cosmological tracking solutions,'' Phys. Rev. D {\bf 59}, 123504 (1999).

\bibitem{leand}
L.~Perivolaropoulos and F.~Skara,``Challenges for $\Lambda$CDM: An update,'' [arXiv:2105.05208 [astro-ph.CO]].

\bibitem{cgm}A. Kamenshchik, U. Moschella, V. Pasquier, ``An alternative to quintessence,'' Phys. Lett. B {\bf 511}, 265 (2001).

\bibitem{41}V. Gorini, A. Kamenshchik, U. Moschella, ``Can the Chaplygin gas be a plausible model for dark energy?,'' Phys. Rev. D {\bf 67}, 063509 (2003).

\bibitem{cgdb1} H.B. Sandvik, M. Tegmark, M. Zaldarriaga, I. Waga, `` The end of unified dark matter?,'' Phys. Rev. D {\bf 69}, 123524 (2004).

\bibitem{cgdb2}R. Bean, O. Dore, ``Are Chaplygin gases serious contenders for the dark energy?,'' Phys. Rev. D {\bf 68}, 023515 (2003).

\bibitem{Zhang1}H. Zhang, Z. H. Zhu, ``Interacting Chaplygin gas," Phys. Rev. D {\bf 73}, 043518 (2006).

\bibitem{Saha0}S. Saha, S. Ghosh, S. Gangopadhyay, ``Interacting Chaplygin gas revisited," Mod. Phys. Lett. A {\bf 32}, 1750109 (2017).

\bibitem{gcg1}M.C. Bento, O. Bertolami, A. A. Sen, ``Generalized Chaplygin gas, accelerated expansion and dark energy matter unification,'' Phys. Rev. D {\bf 66}, 043507 (2002).

\bibitem{gcg2}M.C. Bento, O. Bertolami, A. A. Sen, ``Revival of the unified dark energy-dark matter model?,'' Phys. Rev. D. {\bf 70}, 083519 (2004).

\bibitem{gcgobs1}M. C. Bento, O. Bertolami, A. A. Sen, ``Generalized Chaplygin gas and CMBR constraints,'' Phys. Rev. D {\bf 67}, 063003 (2003).

\bibitem{gcgobs2}M. C. Bento, O. Bertolami, A. A. Sen, ``WMAP Constraints on the Generalized Chaplygin Gas Model,'' Phys. Lett. B {\bf 575}, 172  (2003).

\bibitem{gcgobs3}O. Bertolami, A. A. Sen, S. Sen, P. T. Silva, ``Latest Supernova data in the framework of Generalized Chaplygin Gas model,'' Mon. Not. Roy. Astron. Soc. {\bf 353}, 329 (2004).

\bibitem{gcgobs4}J. S. Alcaniz, D. Jain,  A. Dev, ``High-redshift objects and the generalized Chaplygin gas,'' Phys. Rev. D {\bf 67},  043514 (2003).

\bibitem{gcgobs5}M. C. Bento, O. Bertolami, N. M. C. Santos, A .A. Sen, ``Supernovae constraints on models of dark energy revisited,'' Phys. Rev D {\bf 71}, 063501 (2005).

\bibitem{gcgobs6}T. Barreiro, O. Bertolami, P. Torres, `` WMAP five-year data constraints on the unified model of dark energy and dark matter'' Phys. Rev. D {\bf 78}, 043530 (2008).

\bibitem{ngcg}X. Zhang, F.Q. Wu, J. Zhang, ``New generalized Chaplygin gas as a scheme for unification of dark energy and dark matter,'' JCAP {\bf 0601}, 003 (2006).

\bibitem{ngcg02}K. Liao, Y. Pan, Z. H. Zhu, ``Observational constraints on new generalized Chaplygin gas model'' Res. Astron. Astrophys. {\bf 13}, 159 (2013).

\bibitem{ngcg03}J. Wang, Y. B. Wu, D. Wang, W. Q. Yang,  ``The Extended Analysis on New Generalized Chaplygin Gas,'' Chin. Phys. Lett. {\bf 26}, 089801 (2009).

\bibitem{ngcg04} M. Jamil, ``Interacting New Generalized Chaplygin Gas,'' Int. J. Theor. Phys. {\bf 49}, 62 (2010).

\bibitem{ngcg04a} A. Salehi, M. R. Setare, A. Alaii, ``Reconstructing cosmographic parameters from different
cosmological models: case study. Interacting new generalized Chaplygin gas model,'' Eur. Phys. J. C {\bf 78}, 495 (2018).

\bibitem{ngcg01}F. Salahedin, R. Pazhouhesh, M. Malekjani, ``Cosmological constrains on new generalized Chaplygin gas model, '' Eur. Phys. J. Plus {\bf 135}, 429  (2020).

\bibitem{mamonngcg}A. A. Mamon, V. C. Dubey, K. Bamba, ``Statefinder and $O_{m}$ Diagnostics for New Generalized
Chaplygin Gas Model,'' Universe {\bf 7}, 362 (2021).

\bibitem{jd1} J.D. Barrow, \textquotedblleft Graduated Inflationary
Universe,\textquotedblright\ Phys. Lett. B {\bf 235}, 40 (1990).

\bibitem{jd2} J.D. Barrow and P. Saich, \textquotedblleft The Behavior of intermediate inflationary universes\textquotedblright , Phys. Lett. B {\bf 249}, 406 (1990).

\bibitem{anin1} J.D. Barrow and A. Paliathanasis, \textquotedblleft
Observational Constraints on New Exact Inflationary Scalar-field
Solutions,\textquotedblright\ Phys. Rev.\ D \textbf{94}, 083518 (2016).

\bibitem{anin2} J.D. Barrow and A. Paliathanasis, \textquotedblleft
Reconstructions of the dark-energy equation of state and the inflationary
potential,\textquotedblright\ Gen. Relativ. Grav. \textbf{50}, 82 (2018).

\bibitem{qq} P. Ratra and L. Peebles, \textquotedblleft "Cosmological
consequences of a rolling homogeneous scalar field,\textquotedblright\ Phys. Rev. D {\bf 37}, 3406 (1988).

\bibitem{inf1} A.R. Liddle, \textquotedblleft Chaotic
Inflation,\textquotedblright\ Phys. Lett. B \textbf{129}, 177 (1983).

\bibitem{inf2} P. Parsons and J.D. Barrow, \textquotedblleft Generalised
Scalar Field Potentials and Inflation,\textquotedblright\ Phys.\ Rev.\ D {\bf 51},
6757 (1995).

\bibitem{inf3} A.R Liddle, A. Mazumdar and F.E. Schunck, \textquotedblleft
Assisted inflation\textquotedblright  Phys. Rev. D \textbf{58}, 061301
(1998).

\bibitem{l01} A.R. Liddle, P. Parson and J.D. Barrow, \textquotedblleft
Formalising the Slow-Roll Approximation in Inflation,\textquotedblright\
Phys. Rev. D \textbf{50}, 7222 (1994).

\bibitem{pin2018} Y. Akrami et al. [Planck Collaboration], \textquotedblleft
Planck 2018 results. X. Constraints on inflation,\textquotedblright\ A$\&$A \textbf{641}, A10 (2020).

\bibitem{meq1}E. V. Linder, A. Jenkins, ``Cosmic Structure Growth and Dark Energy,'' Mon. Not. Roy. Astron. Soc. {\bf 346}, 573 (2003).

\bibitem{meq2}A. Bueno belloso, J. Garcia-Bellido, and D. Sapone, ``A parametrization of the growth index of matter perturbations in various Dark Energy models and observational prospects using a Euclid-like survey,'' JCAP {\bf 1110}, 010 (2011).

\bibitem{meq3}L. Amendola, M. Kunz and D. Sapone, ``Measuring the dark side (with weak lensing),'' JCAP, {\bf 0804}, 013 (2008).

\bibitem{meq4} S. Nesseris, ``Matter density perturbations in modified gravity models with arbitrary coupling between matter and geometry,'' Phys. Rev. D {\bf 79}, 044015 (2009).

\bibitem{meq5}S. Nesseris, G. Pantazis, L. Perivolaropoulos, ``Tension and constraints on modified gravity parametrizations of $G_{eff}(z)$ from growth rate and Planck data,'' Phys. Rev. D {\bf 96}, 023542 (2017).

\bibitem{meq6}S. Nesseris, D. Sapone, ``Accuracy of the growth index in the presence of dark energy perturbations,'' Phys. Rev. D {\bf 92}, 023013 (2015).

\bibitem{meq7}S. Nesseris, A. Mazumdar, ``Newton's constant in $f(R,R_{\mu \nu}R^{\mu \nu},\Box R)$ theories of gravity and constraints from BBN,'' Phys. Rev. D {\bf 79}, 104006 (2009).

\bibitem{meq8}S. Tsujikawa, ``Matter density perturbations and effective gravitational constant in modified gravity models of dark energy,'' Phys. Rev. D {\bf 76}, 023514 (2007).

\bibitem{meq9}A. De Felice, S. Tsujikawa, ``$f(R)$ theories,'' Living Rev. Rel. {\bf 13}, 3 (2010).

\bibitem{meq10}A. De Felice, S. Mukohyama, S. Tsujikawa, ``Density perturbations in general modified gravitational theories,'' Phys. Rev. D {\bf 82}, 023524 (2010).

\bibitem{giapr1}H. Steigerwald, J. Bel, C. Marinoni, ``Probing non-standard gravity with the growth index:
a background independent analysis,'' JCAP {\bf 05}, 042 (2014).

\bibitem{giap1}G. Papagiannopoulos, S. Basilakos, A. Paliathanasis, S. Savvidou, and P.C. Stavrinos, ``Finsler-Randers Cosmology:  dynamical analysis and growth of matter perturbations,'' Class. Quantum Grav. {\bf 34}, 225008 (2017).

\bibitem{giap2}W. Khyllep, A. Paliathanasis, J. Dutta, ``Cosmological solutions and growth index of matter perturbations in $f(Q)$gravity,'' Phys. Rev. D {\bf 103}, 103521 (2021).

\bibitem{kahsub}J. C. Bueno Sanchez, L. Perivolaropoulos, ``Evolution of Dark Energy Perturbations in Scalar-Tensor Cosmologies,'' Phys. Rev. D {\bf 81}, 103505 (2010).

\bibitem{dgrf1} Li-Min Wang,  P. J. Steinhardt, ``Cluster abundance constraints on quintessence models,'' Astrophys. J. {\bf 508}, 483 (1998).

\bibitem{dgrf2}E. V. Linder, ``Cosmic growth history and expansion history,'' Phys. Rev. D {\bf 72}, 043529 (2005).

\bibitem{dgrf3}D. Polarski, R. Gannouji, ``On the growth of linear perturbations,” Phys. Lett. B {\bf 660},
439 (2008).

\bibitem{s875}P. Ade et al. (Planck), ``Planck 2013 results. XX. Cosmology from Sunyaev-Zeldovich cluster counts,'' Astron. Astrophys. {\bf 571}, A20
(2014).

\bibitem{s8772}H. Hildebrandt et al., ``KiDS-450: Cosmological parameter constraints from tomographic weak gravitational lensing'' Mon. Not. Roy. Astron. Soc. {\bf 465},
1454 (2017).

\bibitem{s8811}N. Aghanim et al. (Planck), ``Planck 2018 results. VI. Cosmological parameters,'' A$\&$A {\bf 641}, A6 (2020).

\bibitem{refwx}D. Benisty, ``Quantifying the $S_{8}$ tension with the Redshift Space Distortion data set,''  Physics of the Dark Universe {\bf 31}, 100766 (2021).

\bibitem{gsl-a}R. Brustein, `` Generalized second law in cosmology from causal boundary entropy," Phys. Rev. Lett. {\bf 84}, 2072 (2000).

\bibitem{gsl-b} R.M. Wald, ``The thermodynamics of black holes," Living Reviews in Relativity {\bf 4}, 6 (2001).

\bibitem{gsl-c} S. Carlip, ``Black hole thermodynamics," International Journal of Modern Physics D {\bf 23}, 1430023 (2014).

\bibitem{gsl01}S.W. Hawking, ``Particle creation by black holes," Commun. Math. Phys. {\bf 43}, 199 (1975).

\bibitem{gsl02}P. Bin{\'e}truy and A. Helou, ``The apparent universe," Class. Quantum Grav. {\bf 32}, 205006 (2015).

\bibitem{gsl03}A. Helou, "Dynamics of the Cosmological Apparent Horizon: Surface Gravity $\&$ Temperature," arXiv:1502.04235 [gr-qc].

\bibitem{gsl04}S. Saha, "Viaggiu entropy and the generalized second law in a flat FLRW universe," Int. J. Mod. Phys. A {\bf 34}, 1950193 (2019).

\bibitem{gsl05}J.D. Bekenstein, "Black Holes and Entropy," Phys. Rev. D {\bf 7}, 2333 (1973).

\bibitem{gsl06}J.P. Mimoso and D. Pav{\'o}n, "Considerations on the thermal equilibrium between matter and the cosmic horizon," Phys. Rev. D {\bf 94}, 103507 (2016).

\end{thebibliography}
\end{document}